
\def\bs{\bigskip}
\def\ms{\medskip}
\def\np{\vfill\eject}

\def\ni{\noindent}
\def\cl{\centerline}

\def\ref#1#2#3#4{#1\ {\it#2\ }{\bf#3\ }#4\par}
\def\refb#1#2#3{#1\ {\it#2\ }#3\par}
\def\CQG{Class.\ Qu.\ Grav.}
\def\GRG{Gen.\ Rel.\ Grav.}
\def\PR{Phys.\ Rev.}
\def\PRS{Proc.\ R.\ Soc.\ Lond.}

\magnification=\magstep1

\cl{\bf Comment on ``Boundary conditions for the scalar field}
\cl{\bf in the presence of signature change''}
\bs\cl{\bf Sean A. Hayward}
\cl{Department of Physics, Kyoto University, Kyoto 606-01, Japan}
\cl{\tt hayward@murasaki.scphys.kyoto-u.ac.jp}
\bs\ni
{\bf Abstract.}
Fundamental errors exist in the above-mentioned article,
which attempts to justify previous erroneous claims
concerning signature change.
In the simplest example,
the authors' proposed ``solutions'' do not satisfy the relevant equation,
as may be checked by substitution.
These ``solutions''
are also different to the authors' originally proposed ``solutions'',
which also do not satisfy the equation.
The variational equations obtained from the authors' ``actions''
are singular at the change of signature.
The authors' ``distributional field equations'' are manifestly ill defined.

\bs\ni
Several years ago, Dray et al.\ (1991) reported solutions
to the wave equation across a change of signature
which do not satisfy the usual junction conditions,
namely that the momentum fields vanish at the change of signature.
These conditions are well known in quantum cosmology,
where signature-changing solutions are known as real tunnelling solutions
(e.g.\ Gibbons \& Hartle 1990).
The junction conditions were also derived by various purely classical methods
(Hayward 1992, 1993, Kossowski \& Kriele 1993, 1994).
However,
Dray et al.\ (1993) later claimed to have justified their original claims,
and Hellaby \& Dray (1994) went on to claim that
matter conservation is violated at a change of signature.
Some of the mistakes behind these claims have been identified already
(Hayward 1993, 1995), but have not been accepted by the authors concerned.

Nevertheless, it was recently shown that though the ``solutions''
of Dray et al.\ do not satisfy the relevant field equation,
they could be re-interpreted as {\it weak}\/ solutions,
if the equation were written in a certain form (Hayward 1994).
Dray et al.\ (1995) have rejected this re-interpretation
and again attempted to justify their previous claims.
This latest article is considered in the following.
It contains several fundamental errors,
some of which had already been identified in the articles cited above.
It also seriously misrepresents some of these articles.

The authors' own equation numbering and section heading will be used.

\ms\ni{\it Example.}

The authors' first substantive claim, concerning the equation
$$2t\ddot\Phi=\dot\Phi\eqno(1)$$
is that ``The general solution to (1) is''
$$\Phi=\cases{A(-t)^{3/2}+B&($t<0$)\cr Ct^{3/2}+D&($t>0$).\cr}\eqno(2)$$
This is not true.
To check it, simply substitute the proposed solution (2) into the equation (1);
this involves differentiating the function $\Phi$ twice,
but $\Phi$ is not twice differentiable,
unless $A=C=0$ and $B=D$.

The authors then restrict their ``general solution''
by taking $A=-C$ and $B=D$,
for reasons which seem arbitrary;
why restrict the general solution?
Anyway,
their preferred ``solution'' (4) still does not satisfy (1) unless $A=0$,
as is again checked by substitution.
The authors present $A=-C$ and $B=D$ as their ``choice''
and demonstrate their liberal nature by suggesting an alternative ``choice'',
$A=C$ and $B=D$.
But there is no choice;
for (2) to satisfy (1) requires $A=C=0$ and $B=D$.

Moreover, the authors' preferred ``choice'' of matching condition (3),
giving $A=-C$, is the opposite of their preferred ``choice''
in the next two sections of their article,
equations (8) and (25), which correspond to $A=C$.
Their preferred ``choice'' was also $A=C$ in their original article
(Dray et al. 1991).
Given that the authors' professed aim is to justify their original claims,
the confusion seems inexplicable.
Certainly their current article is inconsistent.

The authors end this section by citing the irrelevant example
of a hollow charged conductor in electrostatics.
In that case,
the Maxwell equations have a distributional source---the charge density
being proportional to the normal derivative of the Dirac delta distribution
with support at the dipole layer---which forces discontinuities.
There is no distributional source in equation (1).
Having mentioned distributions, note that the authors' proposed ``solutions''
are clearly written as real functions rather than distributions.

\ms\ni{\it Variational approach.}

The authors' four suggested ``actions'' are not integrals of a Lagrangian form,
but are each the sum of separate integrals over the two domains.
The consequent variational equations
are therefore well defined only away from the junction.
However,
the problem was to make sense of the field equations {\it at}\/ the junction,
where they are singular.
Otherwise, one has not described a {\it change}\/ of signature.
Similar remarks apply to the recent paper of Embacher (1995)
cited by the authors.
Actually, the standard ADM action yields
well defined variational (Lagrange or Hamilton) equations (Hayward 1992).

\ms\ni{\it Distributional wave equation.}

The authors' ``distributional field equation'' is ill defined.
Using (14) and the definitions in the text between (22) and (23),
the ``distributional field equation'' (17) reads explicitly
$$d\left(\Theta^+{*}d\Phi|_{U^+}+\Theta^-{*}d\Phi|_{U^-}\right)=0.$$
This is manifestly ill defined
since it involves multiplication of distributions $\Theta^\pm$ on $M$
by forms which are defined only on part of $M$.
The problem that the authors are trying to avoid is that
the wave equation $d{*}d\Phi=0$ (or ${*}d{*}d\Phi=0$)
is singular at the junction,
since $*$ acting on 1-forms $d\Phi$ is singular,
due to factors of $t^{-1/2}$ or $|t|^{-1/2}$.
The authors have not solved this problem, merely hidden it away in formalism.
This problem has been solved
in terms of a well defined distributional formulation of
the wave equation in the form $d{*}d\Phi=0$ (Hayward 1993).
The usual junction conditions follow.

Also, the equations (21) defining the Hodge dual are inconsistent,
since they imply ${*}{*}1=\hbox{sgn}\,(t)$,
whereas it is well known that ${*}{*}1=1$.

\ms\ni{\it Remarks.}

There are many other criticisms which could be made,
but I will restrict myself to
the authors' principal misrepresentations of my paper
(Hayward 1994; the authors' Ref.~1).
The authors seem to be particularly annoyed that
``Hayward [1] dismisses our approach claiming that it leads to non-unique
solutions and hence ``destroys predictability''. This claim is false.''
Yet I never made such a claim.
The authors ascribe this view to me in five separate places in their paper;
rebutting it seems to be their main concern.
My paper was quite clear on the matter,
but I will repeat the main points as follows.
(i) The ``solutions'' of Dray et al.\ (1991)
do not satisfy their field equations.
(ii) These ``solutions'' can be re-interpreted as {\it weak}\/ solutions,
meaning that
$\Phi$ is re-interpreted as a distribution rather than a real function,
but only if the field equation is written in a certain form.
(iii) There are many more weak solutions than Dray et al.\ suspected;
this is where non-uniqueness and lack of predictability enter.
(iv) There seems to be no preferred way to restore uniqueness,
just arbitrary restrictions.
(v) There was never much reason to look for weak solutions anyway.
To this I will now add:
(vi) in general, such field equations as the Einstein or Klein-Gordon equations
have kinetic terms quadratic in the velocities,
which in a distributional approach
would involve meaningless products of distributions.
Thus such weak solutions will not exist in general.

Note also that the authors' ``general solution'' (2)
cannot be re-interpreted as a weak solution in general, but only if $B=D$.
I certainly never claimed that it could; the authors' claims that
``in [1],
Hayward examines the solutions (2) without imposing any further conditions''
and ``It is these solutions which Hayward claims exhibit non-uniqueness''
are pure fantasy.
I am also mystified by the authors' insistence that they
``have never suggested using solutions of this form''.
I never suggested that they did. I never suggested doing so myself.
Actually, the authors {\it have}\/ now suggested doing so,
by describing (2) as the ``general solution''.

\ms\ni{\it Conclusion.}

In summary:
in different parts of their article,
the authors prefer different ``choices'' of ``solutions''
which are generally inconsistent;
for either ``choice'',
their ``solutions'' do not actually satisfy the relevant equation;
the authors' variational equations, which they do not write explicitly,
are singular;
and their ``distributional field equations'',
which they also do not write explicitly, are manifestly ill defined.

This invalidates the authors' principal contentions,
encapsulated in such righteous pronouncements as
``Contrary to Hayward's claims,
our approach is completely consistent and makes sense of our field equation''
and ``We have shown that Hayward's criticisms of our work our untenable.
Our field equations make sense''.
Unfortunately,
their latest attempts have added further inconsistencies
and still have not made sense of the field equations,
nor of their original ``solutions''.

\bs\ni Acknowledgement:
I discussed an early version of the authors' article with R.W.~Tucker.
(This was not acknowledged by the authors).

\np
\begingroup
\parindent=0pt\everypar={\global\hangindent=20pt\hangafter=1}\par
{\bf References}\par
\ref{Dray T, Manogue C A \& Tucker R W 1991}\GRG{23}{967}
\ref{Dray T, Manogue C A \& Tucker R W 1993}\PR{D48}{2587}
\refb{Dray T, Manogue C A \& Tucker R W 1995}
{Boundary conditions for the scalar field in the presence of signature change}
{(gr-qc/9501034)}
\refb{Embacher F 1995}{Actions for signature change}{(gr-qc/9501004)}
\ref{Gibbons G W \& Hartle J B 1990}\PR{D42}{2458}
\ref{Hayward S A 1992}\CQG9{1851; erratum 2453}
\refb{Hayward S A 1993}{Junction conditions for signature change}
{(gr-qc/9303034)}
\ref{Hayward S A 1994}\CQG{11}{L87}
\refb{Hayward S A 1995}{Comment on ``Failure of standard conservation laws
at a classical change of signature''}{(preprint)}
\ref{Hellaby C \& Dray T 1994}\PR{D49}{5096}
\ref{Kossowski M \& Kriele M 1993}\CQG{10}{2363}
\ref{Kossowski M \& Kriele M 1994}\PRS{A444}{297}
\endgroup
\bye